\documentclass[conference]{IEEEtran}
\IEEEoverridecommandlockouts

\usepackage{subcaption}

\usepackage[ruled,linesnumbered]{algorithm2e}
\usepackage{amsmath,amssymb,amsfonts}
\usepackage{graphicx}
\usepackage{soul,color}
\usepackage{authblk}

\setstcolor{red}

\begin{document}


\title{Federated Learning based Hierarchical 3D Indoor Localization}







\author[1]{Yaya Etiabi}
\author[2]{Wafa Njima}
\author[1]{El Mehdi Amhoud}
\affil[1]{School of Computer Science, Mohammed VI Polytechnic University, Benguerir, Morocco}
\affil[2]{ISEP, Institut Sup\'{e}rieur d'Electronique de Paris, 75006 Paris, France  \authorcr Emails: {\{yaya.etiabi, elmehdi.amhoud\}@um6p.ma, wafa.njima@isep.fr}}



\maketitle

\begin{abstract}

The proliferation of connected devices in indoor environments opens the floor to a myriad of indoor applications with positioning services as key enablers. However, as privacy issues and resource constraints arise, it becomes more challenging to design accurate positioning systems as required by most applications. To overcome the latter challenges, we present in this paper, a federated learning (FL) framework for hierarchical 3D indoor localization using a deep neural network. Indeed, we firstly shed light on the prominence of exploiting the hierarchy between floors and buildings in a multi-building and multi-floor indoor environment. Then, we propose an FL framework to train the designed hierarchical model. The performance evaluation shows that by adopting a hierarchical learning scheme, we can improve the localization accuracy by up to 24.06\% compared to the non-hierarchical approach. We also obtain a building and floor prediction accuracy of 99.90\% and 94.87\% respectively. With the proposed FL framework, we can achieve a near-performance characteristic as of the central training with an increase of only 7.69\% in the localization error. Moreover, the conducted scalability study reveals that the FL system accuracy is improved when more devices join the training.

\end{abstract}

\section{Introduction}
Location-based services (LBS) are critical components for Internet of Things (IoT) applications since most IoT devices operate on location-dependent data. Indeed, most IoT applications such as emergency services, asset tracking, logistics planning, e-marketing, and social networking critically exploit location information to operate properly. With the unprecedented expansion of IoT networks, more efforts are needed to design suitable and accurate localization systems exploiting the new possibilities brought by new-generation communication systems, in contrast to the global positioning system (GPS). This latter is energy-intensive in addition to its high deployment cost and the environmental challenges it faces, especially in indoor environments. Compared to outdoor positioning \cite{Etiabi2022SpreadingFA}, indoor localization is indeed the most challenging localization problem most often addressed using wireless signal parameters such as time of arrival (ToA), time difference of arrival (TDoA), angle of arrival (AoA) and received signal strength indicator (RSSI) \cite{etiabi2020}. RSSI is the most widely used signal property for indoor as well as outdoor localization due to its low cost and high availability, as enlightened in \cite{Jouhari2022ASO}. Traditional indoor localization systems are based on geometric methods (i.e. trilateration, multilateration, and triangulation) and fingerprint-based methods. However, geometric methods are heavily affected by multipath propagation effects and RSSI fingerprinting suffers from an inherent problem caused by the very indoor dynamic environment and unstable wireless devices. This makes the localization accuracy degrades abruptly over time, requiring a high calibration effort for fingerprint  collection \cite{Njima2020}. 
Additionally, with the increase in the number of connected devices, the fingerprinting method faces many channel impairments such as interference between signals and non-line-of-sight (NLOS) propagation due to the presence of many reflecting surfaces in indoor environments, leading to large localization errors.

As a result of the difficulty in developing robust models that capture these indoor channel impairments, researchers are turning towards data-based localization using machine learning (ML) techniques, which do not require empirical models but instead rely on offline constructed datasets that capture variations in indoor environments.
Indeed, as demonstrated in \cite{Marwa_semiSupervised,Marwa_conv}, ML algorithms, especially deep neural networks (DNNs) are employed to address the aforementioned drawbacks of traditional approaches. It is observable from these studies that ML is a very promising technology for IoT localization since it enables robust and scalable localization systems with enhanced accuracy and a relatively low online complexity \cite{scalableDNNLocIndoor2017}. 
Nonetheless, DNN architectures proposed in the aforementioned works do not take into account the hierarchical nature of the localization task in multi-building and multi-floor indoor environments which we shed light on in this paper. Authors in \cite{RNNhierachical2021} and \cite{auxiliary2022} have drawn attention to the importance of considering the hierarchical aspect of this indoor localization task using DNN algorithms.

However, ML-based localization solutions entail the collection of data from IoT devices into a central server, which results in a lot of data exchange with the server, privacy concerns, and a high reliance on the server. Note that in addition to the high bandwidth requirement, this assumes the server to be trustworthy. This may lead to issues in practical considerations where data are distributed across IoT devices in a privacy-preserving objective. Consequently, federated learning (FL) was developed to conserve bandwidth while protecting the privacy of users' data.
Indeed, FL is becoming more attractive for localization problems with its great advantage of privacy by default and bandwidth optimization, especially in wireless communications involving resource-constrained IoT devices. In \cite{FedLoc2020}, authors have shown the prominence of using such an algorithm for indoor localization but to date, only a few research works have really tackled this problem. Authors in \cite{Etiabi2022FederatedDB} have solved the indoor localization problem leveraging a more communication-friendly FL scenario called \textit{federated distillation} where only
the model outputs whose dimensions are commonly much smaller than the model size are exchanged.
Although, the proposed method was only restricted to location estimation, and no information about either the building or the floor was provided.

The main focus of this work is to first design a hierarchical learning scheme for the joint building, floor, and precise 2D coordinates prediction in multi-building \& multi-floor indoor environments. Then, regarding the resource constraints of IoT devices, we propose a federated learning framework to train the proposed hierarchical model, yielding a communication-efficient collaborative and privacy-preserving indoor localization solution.
The main contributions of this work can be summarized as follows:
%
%
\begin{itemize}
    \item We develop a novel 3D indoor
positioning system with a new DNN architecture incorporating the hierarchical nature of indoor localization tasks in  multi-building and multi-floor indoor environments. A single hierarchical DNN model is proposed to jointly predict the buildings, the floors, and the precise 2D coordinates of the users.
 Then, we validate the results with a publicly available experimental indoor localization dataset which makes our solution more realistic for indoor IoT applications.
    \item We propose a federated training of the proposed architecture to preserve IoT devices location data privacy and save the bandwidth of the wireless infrastructure. Consequently, we provide a collaborative, bandwidth-optimization, and  privacy-preserving indoor localization solution for IoT applications. 
    \item Being aware of the exponential growth of IoT networks, we investigate the scalability of the proposed FL framework and we provide an analysis of different wireless transmissions involved in the overall learning process.
\end{itemize}
The remainder of this paper is organized as follows:
Section~\ref{sec:hdnn} introduces the hierarchical learning scheme for our localization problem while Section~\ref{sec:hdnn} depicts the proposed FL framework. In Section~\ref{sec:performance-eval}, we go through the performance evaluation and analysis of our framework before concluding in Section~\ref{sec:conclusion}.


\section{System model and Problem formulation}
\label{sec:hdnn}


In a multi-building and multi-floor indoor environment, the position of the target can be recursively obtained in a hierarchical manner starting with the building identification followed by the floor identification, and finally the fine-grained location of the target. Thus, we can expect to determine the position of the target with more precision. Let $\mathcal{B} = \left\{ b\right\}_{b=1,2,3,..,\mathcal{N}_b}$, \mbox{$\mathcal{F} = \left\{ f\right\}_{f=1,2,3,..,\mathcal{N}_f}$}, and \mbox{$\mathcal{L} = \left\{ (x_l, y_l)\right\}_{x_0,y_0\leq x_l,y_l\leq x_0+L,y_0+W}$} be respectively the sets of buildings, floors, and locations constituting the indoor environment. $L$ and $W$ correspond respectively to the maximum length and the maximum width of the floors. Note that the set of locations $\mathcal{L}$ constitutes a continuous space with $x_0<x_l<x_0+L$ and $y_0<y_l<y_0+W$ where $(x_0, y_0)$ are the reference coordinates.
Let $\mathcal{R}_l$ be the set of RSSI received from the set of detected access points (APs) at location $l$ denoted as $AP_l$. $\mathcal{R}_f$ is the set of RSSI received from APs detected in floor $f$ denoted as $AP_f$. $\mathcal{R}_b$ is the set of RSSI received from APs detected inside building $b$ denoted as $AP_b$. The previous parameters are defined as follows:
\begin{equation}
    \resizebox{.9\hsize}{!}{$\left\{\begin{matrix}
 \mathcal{R}_l&=&\{RSSI_n^{l}\},& AP_l &=& \{AP_n\}, & n=1,2,3,...,M_l \\
\mathcal{R}_f&=&\bigcup_{l}  \mathcal{R}_l,& AP_f &=& \bigcup_{l} AP_l,  & (x_l,y_l)\in \mathcal{F}_f \\
\mathcal{R}_b&=&\bigcup_{f}  \mathcal{R}_f,& AP_b &=& \bigcup_{f} AP_f,  & f\in \mathcal{B}_b  \\
\end{matrix}\right.$},
\end{equation}
where $M_l$ is the number of APs detected at location $l$.
The hierarchical nature of the indoor environment can be used to reduce the search space during a target localization. 
To localize a target, RSSI measurements will be collected in order to determine its position $\theta$, with $\theta =\left(b^*, f^*, l^*=[x_{\theta}, y_{\theta}]\right)$.
The set of RSSI received at such position is  $\mathcal{R}_{\theta}$ with the corresponding set $AP_{\theta}$.

Since the similarities of $\mathcal{R}_{\theta}$ with those of the buildings ($\mathcal{R}_{b}, b\in \mathcal{B}$) allow a clearer distinction, we can reduce the search space by switching from the entire indoor environment to a building chosen based on the best similarity measure. Indeed, to evaluate the similarity, we consider the Euclidean distance, and  accordingly, we denote by $d\left(\mathcal{R}_{\theta}, \mathcal{B}_b\right)$ the mean distance between the measurements in $\mathcal{R}_{\theta}$ with all the measurements $\mathcal{R}_b \in \mathcal{B}_b$ which is defined as follows:
\begin{equation}
    d\left(\mathcal{R}_{\theta}, \mathcal{B}_b\right) = 
    \frac{1}{\left| \mathcal{R}_b\right|}\sum_{n=1}^{\left| \mathcal{R}_b\right|} \left\| \mathcal{R}_{\theta} - \mathcal{R}_{b}^n\right\|_2 .
\end{equation}

Based on calculated distances, the corresponding building is given by $b^* = \arg\min_b \{d\left(\mathcal{R}_{\theta}, \mathcal{B}_b\right)\}$.
The next search space reduction is floor identification. Similarly, the floor of the target is determined by the following equation:
\begin{equation}
    f^* = \arg\min_{f}
    \frac{1}{\left| \mathcal{R}_f\right|}\sum_{n=1}^{\left| \mathcal{R}_f\right|} \left\| \mathcal{R}_{\theta} - \mathcal{R}_{f}^n\right\|_2, \\\text{subject to }\mathcal{R}_{f} \in \mathcal{R}_{b^*}
\end{equation}

At this stage, the building and the floor of the target are estimated and we can proceed to the prediction of its fine-grained location $(x_\theta, y_\theta)$. We consider a regression model to fit the RSSI measurements of a given floor in a given building to the corresponding locations as follows:\mbox{ $\mathbf l = \mathbf{X}_{(b,f)}\mathbf{\beta}_{(b,f)} + \mathbf{\epsilon}_{(b,f)},$}
where $\mathbf{l} = (x_l,y_l)$, $\mathbf{X}_{(b,f)}=  \mathcal{R}_{f} \in \mathcal{R}_{b}$ is the RSSI observations on the floor $f$ belonging the building $b$, $\mathbf{\beta}_{(b,f)}$ and $\mathbf{\epsilon}_{(b,f)}$ are the corresponding regression model parameters.
The predicted target location $\mathbf{l}^*$ is then determined by \mbox{$\mathbf{l}^* =\mathcal{R}_{\theta}\mathbf{\beta}_{(b^*,f^*)} + \mathbf{\epsilon}_{(b^*,f^*)}.$}
Finally, we obtain the full position estimate of the target $\theta^* =\left(b^*, f^*, \mathbf{l}^*=[x_{\theta^*}, y_{\theta^*}]\right). $

As it can be noticed, this hierarchical approach is too complex to be implemented with empirical approaches since, in addition to the two consecutive KNN-like methods for building and floor identification, a linear regression model is needed for each floor, leading to  highly complex solution undergoing the curse of dimensionality, hence the need for a data-driven neural network approximator.
Accordingly, we design a deep neural network (DNN) that is expected to automatically extract the hierarchical properties in the RSSI dataset and predicted at once the 3D localization of any target device in the indoor environment. Moreover, in practical considerations, regarding the privacy issue and the bandwidth limitation that the crowdsourcing of RSSI arises, we propose  a federated learning framework to collaboratively train the designed DNN using RSSI data distributed across IoT devices.

\section{Federated learning framework for indoor positioning}
\label{sec:floc}
\subsection{Data preprocessing}
We use the dataset presented in \cite{ujiindoorloc}
which is constituted of RSSI measurements from different APs deployed in a campus indoor environment.
Since not all the APs are in range during the measurements, missing RSSI readings are set to 100 dBm. 
In our preprocessing, we first dropped all the APs which have not been captured during all the measurements campaign, i.e the columns with all values missing. This has led to a fingerprint dimension of 465 APs compared to the original dataset with 520 APs.
To further reduce the fingerprint dimension, we define a threshold of visibility of APs, $\tau =0.98$ i.e., if an AP is not captured in more than 98{\%} of measurements, its contribution to the richness of the dataset is not significant and thus can be removed. At the end of this step, we keep only 248 APs.

Once APs selected, we deal with missing RSSIs replacing them with a more convenient value which is chosen as the overall minimum RSSI reading diminished by 1 dBm noted $min_{RSSI}$, resulting in $min_{RSSI}$= -105 dBm (i.e, -104 dBm-1 dBm). Then, we adopt the technique detailed in \cite{powed}, where RSSI measurements are converted to a powered representation 
defined as follows:
\begin{equation}
    powed\left(RSSI_i\right) = \left(\frac{RSSI_i - min_{RSSI}}{-min_{RSSI}}\right)^\beta,
\end{equation}
where $\beta$ is set to the mathematical exponent $e$. This step results in a positive and normalized representation of the data which is a good fit to enhance the performance of the DNN.

\subsection{Proposed DNN architecture and federated training}
\begin{figure}[!t]
     \centering
     \begin{subfigure}{\columnwidth}
         \centering
         \includegraphics[scale=0.2]{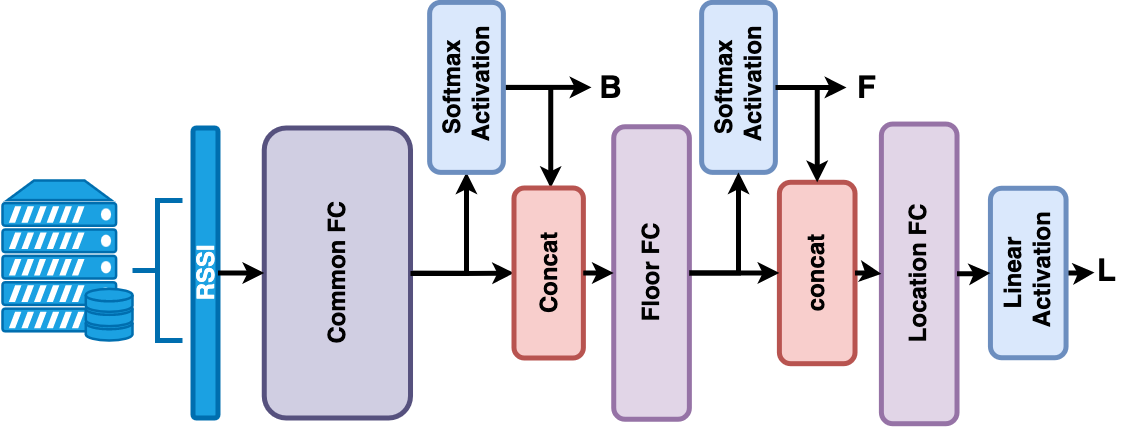}
         \caption{Proposed hierarchical MLP model (H-MLP)}
         \label{fig:h-sysm}
     \end{subfigure}
     \vfill
     \begin{subfigure}{\columnwidth}
         \centering
         \includegraphics[scale=0.2]{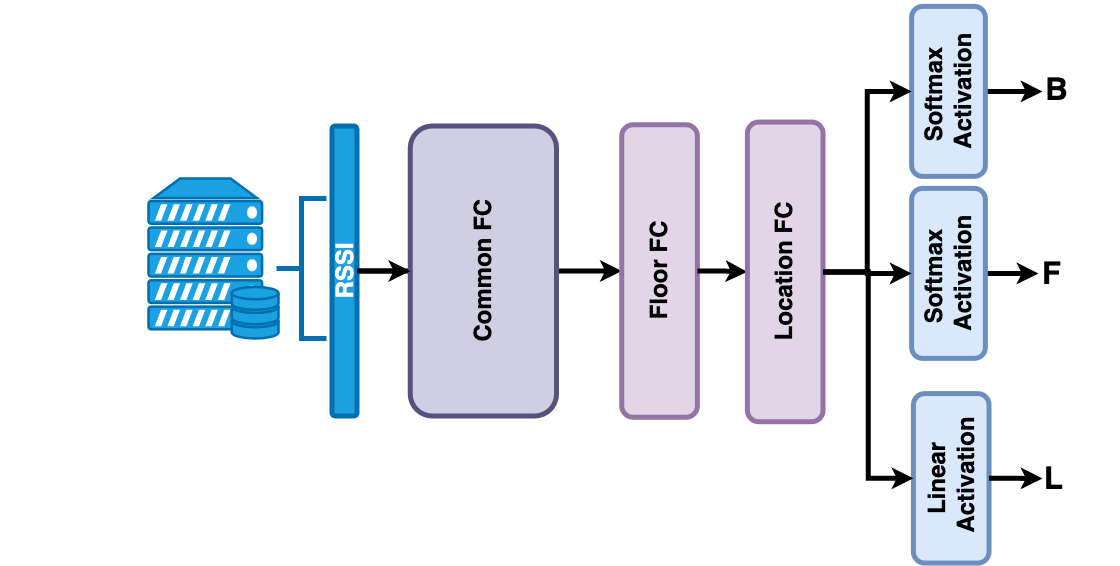}
         \caption{Compacted benchmark MLP  model}
         \label{fig:b-sysm}
     \end{subfigure}
        \caption{3D indoor localization system model.}
        \label{fig:sysm}
\end{figure}

As depicted in Fig.~\ref{fig:h-sysm}, we design a deep learning model namely a multi-layer perceptron (MLP) taking the RSSI as a single input and outputting the full coordinates namely the building, the floor, and the fine-grained 2D location, while exploiting the hierarchical nature of the indoor positioning. We coined our model H-MLP, for hierarchical MLP. Note that this single network is configured to function as a multitask network since it is simultaneously performing two classification tasks (building and floor identification) and a regression one (location coordinates estimation).
The DNN can be represented by an approximation function $F_{\mathbf{W}}(\cdot)$ parameterized by $\mathbf{W}$ representing the weights  of the network. Thus, any position $\theta $ can be estimated by $\hat{\theta} = F_{\mathbf{W}} \left( \mathcal{R}_{\theta} \right)$.
The training of the model which aims to determine the optimal weights $\mathbf{W}^*$ is done through the minimization of the global loss which encompasses three distinct losses corresponding each to a task.
The global loss is thus given by:
\begin{equation}
    L_{\mathbf{W}}(\hat\theta, \theta) = \frac{1}{\sum_{u \in\{b,f,l\}} \alpha_u} \sum_{u \in\{b,f,l\}} \alpha_u L_{\mathbf{W}}^u(\hat\theta_u, \theta_u),
\end{equation}
where $\alpha_u$ are the weights representing the contribution of each loss to the global loss.

To exhibit the hierarchical nature of indoor localization and the importance of exploiting it, we derived a second DNN architecture that directly predicts all the outputs constituted by the building ID, the floor ID, and the 2D location (x,y), by compacting all the layers as shown in Fig.~\ref{fig:b-sysm}. 

Our objective is to train the DNN model previously defined using different RSSI data collected by IoT devices deployed over a wireless network. Without data sharing with a central server for privacy and bandwidth preservation, we turn towards an iterative and collaborative training process called federated learning, where each IoT device participates in the training of the global model using its local dataset, and the whole collaboration is federated by a parameter server. Indeed, the parameter server is in charge of defining the global model and selecting participants also called clients at each training iteration or communication round.
The whole process of the federated model training can be described by Algorithm~\ref{alg:algo}.
\SetKwComment{Comment}{/* }{ */}
\RestyleAlgo{ruled}
\begin{algorithm}
\SetAlgoLined
  \caption{Federated Learning for Localization}
  \label{alg:algo}
  \SetKwInOut{Input}{Inputs}
  \SetKwInOut{Output}{Outputs}
  \Input{$\{\mathcal{D}_c\}$ \Comment*[l]{Clients' RSSI Datasets}}
  \Input{$\{\alpha_u\}$ \Comment*[l]{Outputs' weights}}
  
  \Output{$\{W\}$ \Comment*[l]{Trained model parameters}}
  \SetKwProg{ServerInit}{ServerInit}{}{}
   \ServerInit{()}{
  Set $W_0$,  and $ r \gets 0$ \;
  
  }
  \While{not converged and $r < max\_com\_rounds$}{
  server broadcasts $W_r$\;
  
  \ForEach{Client $c \in \{1,2, \dots,C\}$ }{
  \Comment{In parallel}
  \SetKwProg{ClientLocalTraining}{ClientLocalTraining}{}{}
   \ClientLocalTraining{$(\mathcal{D}_c, \alpha_u, W_r$)}{
    
    Update local model with $W_r$ \;
    Train local model using
    $
    \mathbf{W}_{r,k+1}^c = \mathbf{W}_{r,k}^c - \mu\nabla F_{\mathbf{W}_{r,k}^c}(\mathcal{B}_c)
    $ \;
     Client uploads $W_r^c$ \;
  }

  }
  \SetKwProg{ServerGlobalUpdate}{ServerGlobalUpdate}{}{}
  \ServerGlobalUpdate{$(\{W_r^c, \left|\mathcal{D}_c\right|\})$}{
  Update global model using
   $\mathbf{W}_{r+1} = \frac{1}{\sum_c \left| \mathcal{D}_c \right|} \sum_c  \left| \mathcal{D}_c \right| \mathbf{W}_{r}^c $ \;
  }
  
  $ r \gets r +1$ \;
  
  }
  \end{algorithm}

\subsection{Wireless communication network}
In this part, we focus on the communication aspect of the federated learning system. Regarding the nature of exchanges between clients and the server, the downlink and the uplink transmissions are treated separately.
\subsubsection{Downlink communication}
On the downlink, the global model is broadcast to all the clients by the server, making the communication less constrained than on the uplink.
Indeed, in the $r^{th}$ communication round, the server broadcasts to all the clients in the downlink, so that the signal received by a client $c$ from the server is given by $\mathbf{y}_r^c= \mathbf{g}_r^{c}\mathbf{x}_r+\mathbf{z}_r^c$,
where $\mathbf{x}_r$ is the $T_D\times1$ transmitted signal by the server with $T_D$ the number of downlink channels used, $\mathbf{g}_r^{c}$ is the quasi-static fading channel from the server to the client c, and
$\mathbf{z}_r^c$ is $T_D\times1$ noise vector with independent and identically distributed (i.i.d.) Gaussian entries. Note that the server is subject to the power
constraint $E\left[\left\|\mathbf{x}_{r}\right\|_{2}^{2}\right] / T_{D} \leq \mathcal{P}_{D}$ \cite{com}.
With such assumptions, and considering Shannon capacity \cite{shannon}, the number of broadcast bits by the server to all clients during the $r^{th}$ communication round under digital transmission in the downlink is given by $B_{D, r}=\min _{c}\left(T_{D} \log _{2}\left(1+\left|\mathbf{g}_r^{c}\right|^{2} \mathcal{P}_{D}\right)\right)$,
where $\left|\mathbf{g}_r^{c}\right|^{2}$ is the server-to-client channel gain with the server assumed to have the knowledge of the channel $\mathbf{g}_r^{c}$. 
As a result, the server cannot transmit to clients a model whose bit size of the parameters is larger than this downlink capacity $B_{D, r}$, at least without any compression scheme.
\subsubsection{Uplink communication}
On the uplink, the FL clients share a Gaussian multiple-access channel whose equation is given by $\mathbf{y}_{r}=\sum_{c=1}^{C} \mathbf{h}_{r}^{c} \mathbf{x}_{r}^{c}+\mathbf{z}_{r}$,
where $\mathbf{y}_{r}$ is the signal received at the server, $\mathbf{x}_r^c$ is the $T_U\times1$ signal transmitted to the server by client $c$ with $T_U$ the number of uplink channels used, $\mathbf{h}_r^{c}$ is the quasi-static fading channel from the client $c$ to the server, and
$\mathbf{z}_r$ is $T_U\times1$ noise vector with i.i.d. $N (0, 1)$ entries. Note that the FL clients are subject each to the power constraint $E\left[\left\|\mathbf{x}_{r}^c\right\|_{2}^{2}\right] / T_{U} \leq \mathcal{P}_{U}$.
In a conventional digital system, as shown in \cite{com}, the uplink capacity of the channel is shared evenly among the clients, resulting in bandwidth limitations that limit the number of bits that may be transmitted by each client. Indeed based on Shannon's capacity\cite{shannon}, the maximum amount of bits per transmission for any client $c$ is provided by ${B}^{c}_{U,r}=\frac{T_U}{ C} \log _{2}(1+C \left | \mathbf{h}_r^c \right |^2 \mathcal{P}_U)$,
where $\left|\mathbf{h}_{r}^{c}\right|^{2}$ is the client $c$ to  server channel gain.

\section{performance evaluation}
\label{sec:performance-eval}

We use the UJIIndoorLoc database reported in \cite{ujiindoorloc} to assess the performance of our model. UJIIndoorLoc database contains two datasets: a training set containing 19937 RSSI recordings and a validation set with 1111 recordings taken 4 months after the training set, in order to assimilate a real world use case.
For fair comparison with the state-of-art algorithms using the same database, we divided the training data into training set and a test set with a ratio of 90:10 to train and test our model, and the validation set is used to measure the performance of our model with a benchmark analysis.
The performance of our model is assessed through the building prediction accuracy (B-ACC), the floor prediction accuracy (F-ACC), the success rate also referred to as global accuracy (ACC), and the mean distance error (MDE).
We can distinguish two types of MDE in our evaluation: The 2D-MDE which refers to the mean localization error when the building and the floor are correctly predicted, and can be defined as follows:
\begin{equation}
\resizebox{.9\hsize}{!}{$\text{2D-MDE}=\frac{1}{N} \sum_{i=1}^{N}\left [   \delta \left(\mathcal{B}_i, \hat{ \mathcal{B}_i}\right)\cdot \delta \left( \mathcal{F}_i, \hat{ \mathcal{F}_i}\right)\cdot\sqrt{\left ( x_i-\hat{x}_i \right )^2+\left ( y_i-\hat{y}_i \right )^2}\ \right ]$},
\end{equation}
where $\delta \left(a, b\right) = 1$ if $a=b $, $0$ otherwise.
The 3D-MDE which is the usual MDE representing the global localization error and is defined as: 
 \begin{equation}
   \text{3D-MDE}= \frac{1}{N} \sum_{i=1}^{N} \sqrt{\left ( x_i-\hat{x}_i \right )^2+\left ( y_i-\hat{y}_i \right )^2},
 \end{equation}
where $(x, y)$ and $(\hat{x}, \hat{y})$ are respectively the ground truth and the predicted coordinates.
\subsection{Evaluation of the H-MLP model}
\begin{table}[!t]
    \centering
    \caption{Global model parameters following the configuration depicted in Fig.~\ref{fig:sysm}.}
    \resizebox{0.5\textwidth}{!}{\begin{tabular}{l|l}
    \hline
       \textbf{Common layers } & \textbf{Building layer} \\
       \hline
      \begin{tabular}{ll}
          hidden layers & 256-128 \\
dropout layer & 0.3 \\
activation functions & ReLU\\
Batch normalization layer \\
      \end{tabular} &
      \begin{tabular}{ll}
           output activation & softmax \\
output weight $\alpha_b$ & 0.1 \\
      \end{tabular} \\
       \hline
       \hline
       \textbf{Floor layer} & \textbf{Location layer}\\
      \hline
      \begin{tabular}{ll}
           hidden layer & 128 \\
        dropout layer & 0.1 \\
        Batch normalization layer \\
        output activation & softmax \\
        output weight $\alpha_f$ & 0.3 \\
      \end{tabular} &
      \begin{tabular}{ll}
          hidden layer & 128 \\
        dropout layer & 0.1 \\
        Batch normalization layer \\
        output activation & linear \\
        output weight $\alpha_l$ & 0.6 \\
      \end{tabular}
    \end{tabular}}
    \label{tab:config}
\end{table}

\begin{figure}[!t]
    \centering
    \includegraphics[scale=0.8]{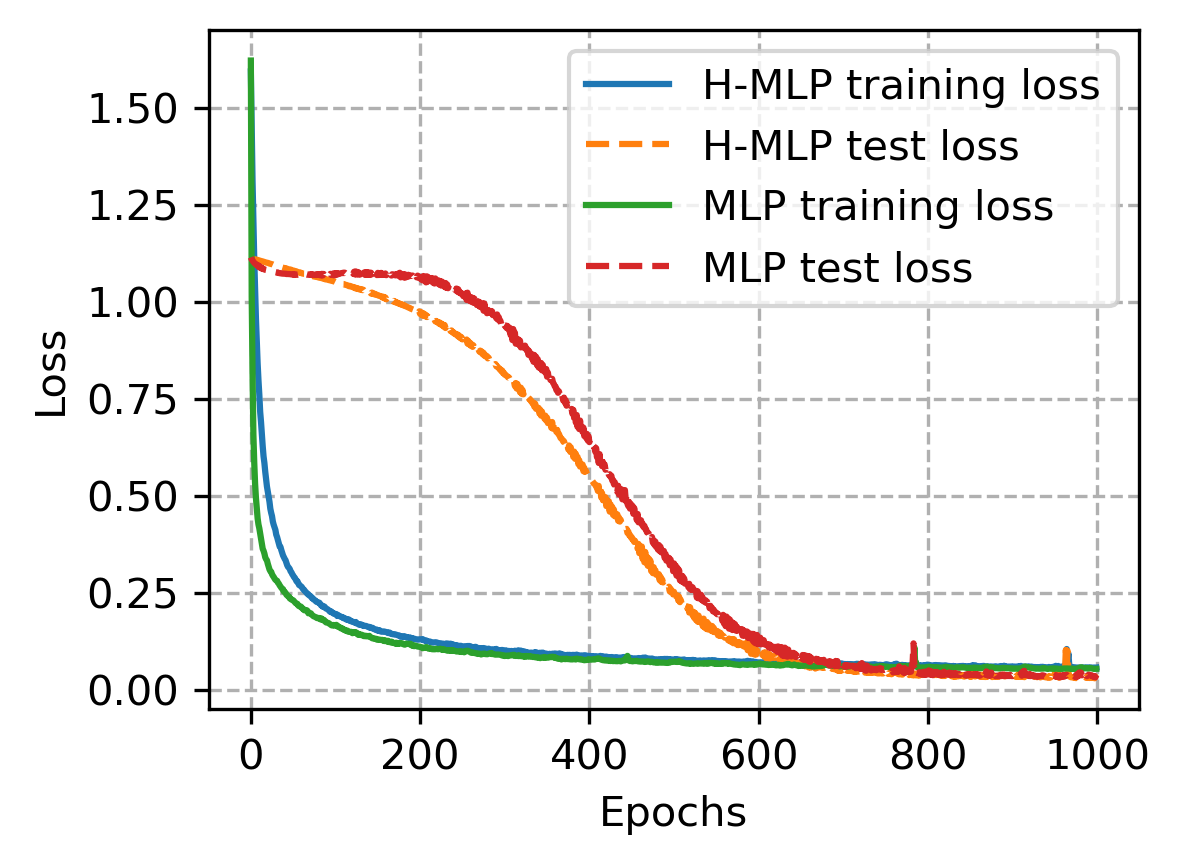}
    \caption{Model training performance.}
    \label{fig:mlp_loss}
\end{figure}

\begin{figure*}[!t]
    \centering
    \begin{subfigure}[b]{0.33\textwidth}
        \centering
        \includegraphics[width=\textwidth,height=0.2\textheight]{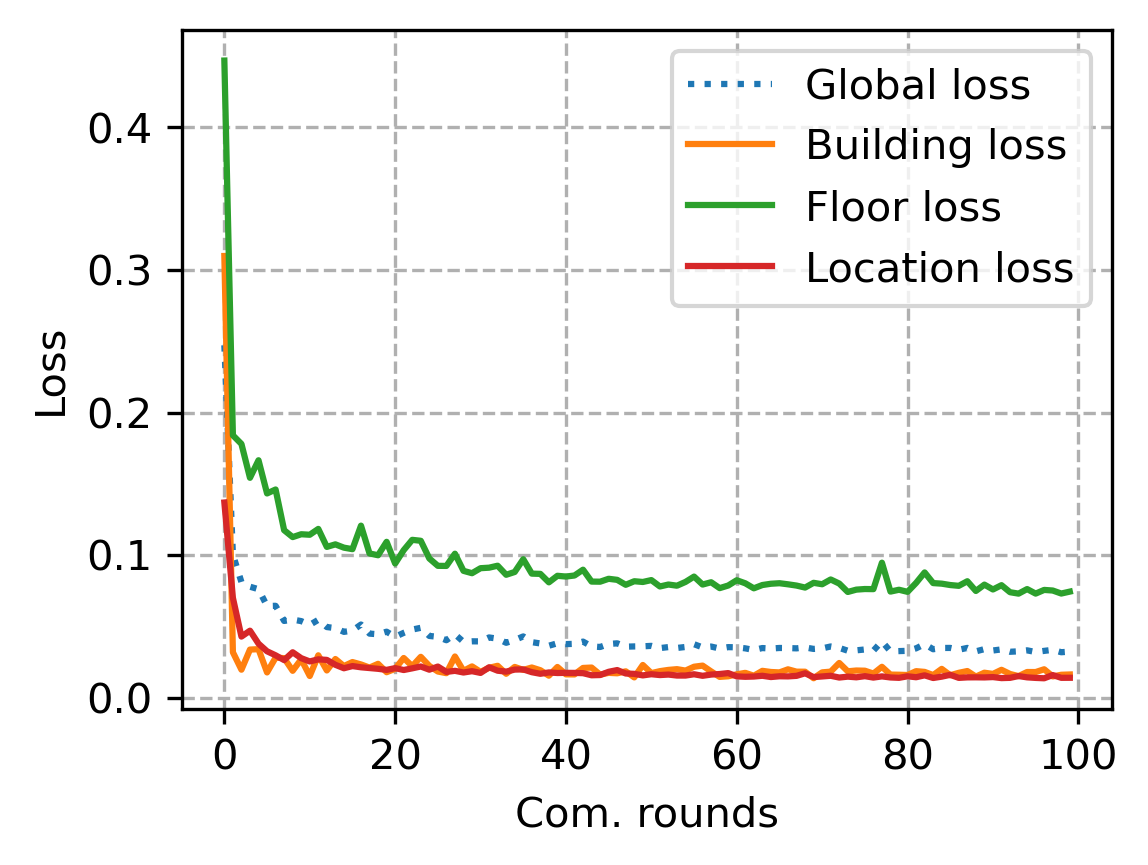}
        \caption{Federated training performance.}
        \label{fig:fl-loss}
    \end{subfigure}%
    \hfill
    \begin{subfigure}[b]{0.33\textwidth}
        \centering
        \includegraphics[width=\textwidth,height=0.2\textheight]{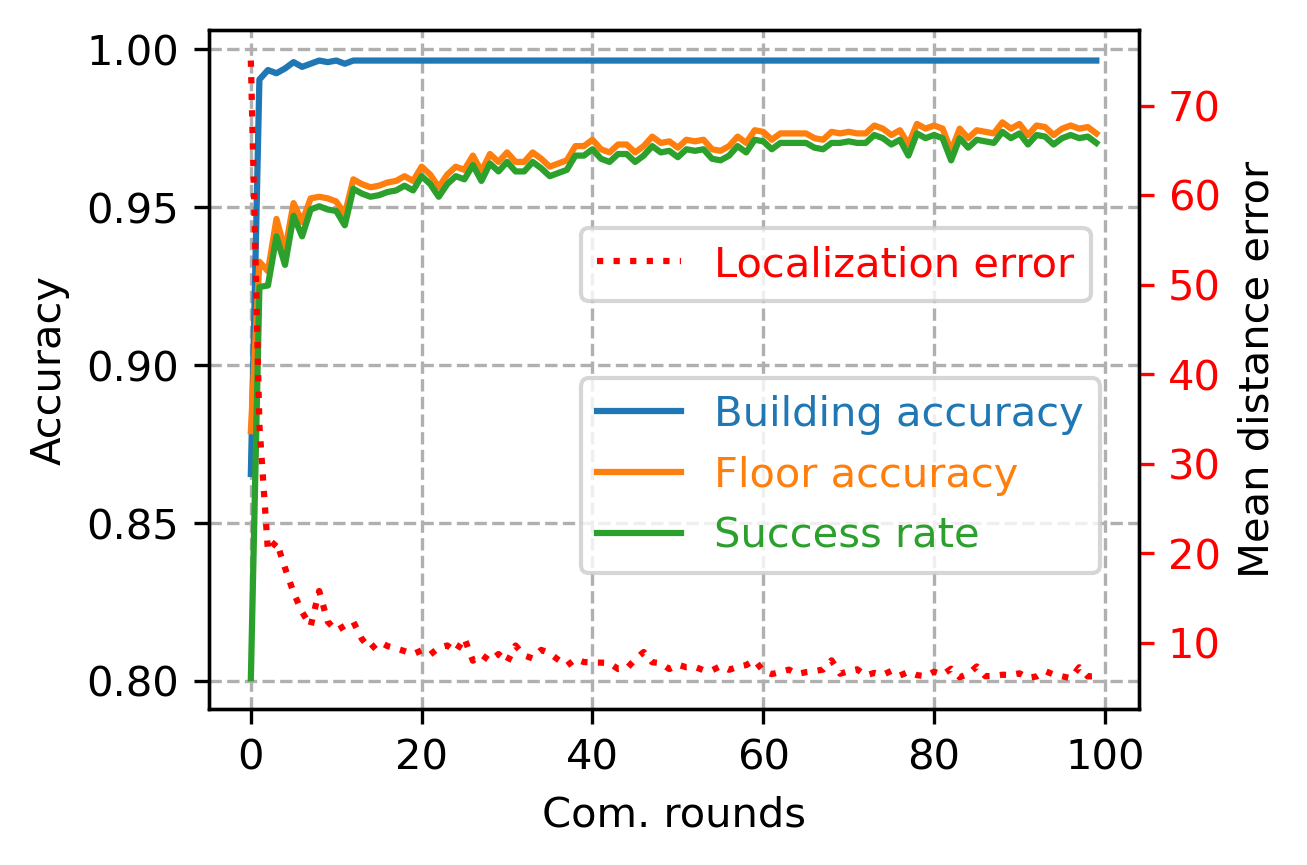}
        \caption{FL system evaluation.}
        \label{fig:fl-accuracy}
    \end{subfigure}
    \hfill
    \begin{subfigure}[b]{0.33\textwidth}
        \centering
        \includegraphics[width=\textwidth,height=0.2\textheight]{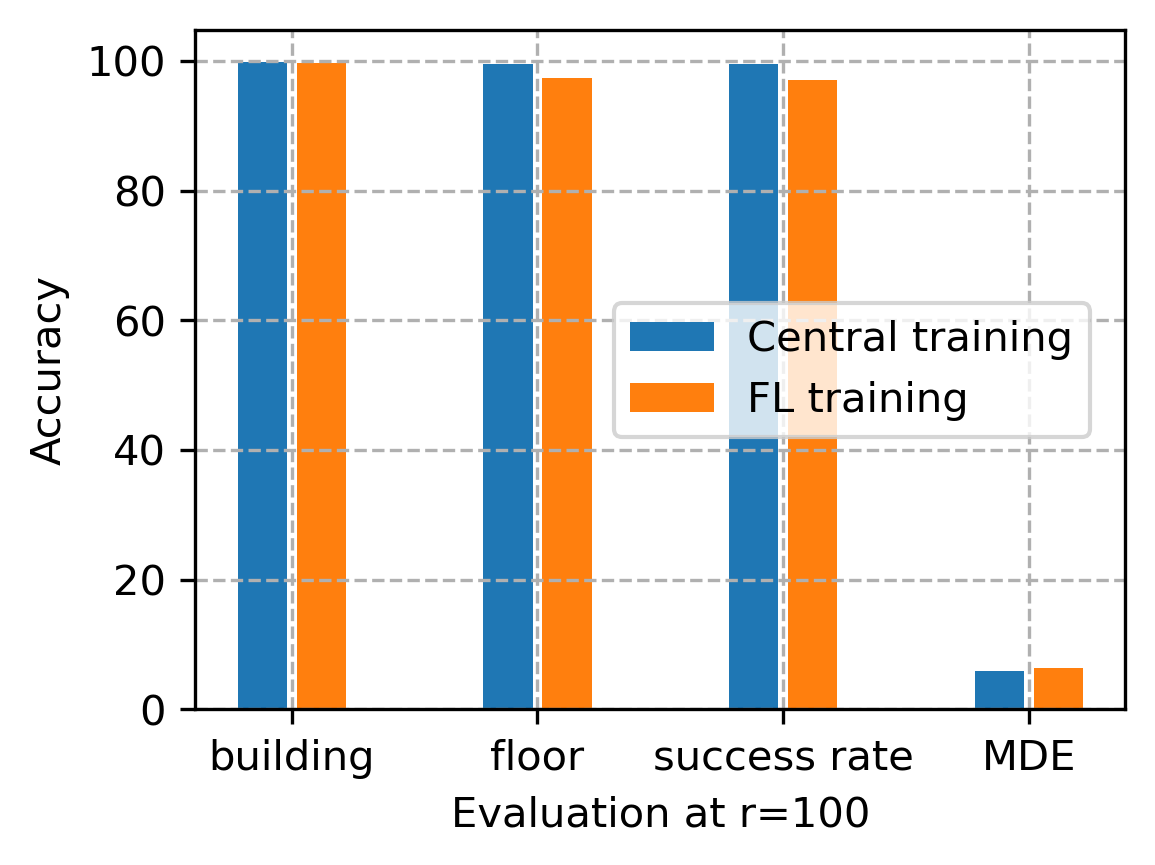}
        \caption{Comparison with centralized learning.}
        \label{fig:fl-central}
    \end{subfigure}
    \caption{FL framework performance evaluation.}
    \label{fig:fl-performance}
\end{figure*}
The global configuration of our model is depicted in TABLE~\ref{tab:config}. Note that for fair comparison, we use the same configuration for our H-MLP as well as for the MLP model.
We ran the model for 1000 epochs with a batch size of 64 using \textit{Adam} optimizer with a learning rate of $5\times10^{-4}$ and the result is shown in Fig.~\ref{fig:mlp_loss}. We can observe that our proposed architecture (H-MLP), as well as the baseline model (MLP), converges as the training approximates 600 epochs, while the evaluation showed that H-MLP improves the localization system performance. As shown in TABLE~\ref{tab:mlp}, H-MLP improves the localization accuracy by up to 24.06$\%$ in comparison to the baseline MLP model.

Moreover, we performed a benchmark analysis of our model by comparing it with state-of-the-art deep learning methods using the same dataset as presented in TABLE~\ref{tab:bcmk}. It can be seen that our model provides a better trade-off between building prediction accuracy, floor prediction accuracy, and localization error. Our method outperforms the state-of-the-art approaches including Scalable DNN\cite{scalableDNNLocIndoor2017} and CNNLoc\cite{CNNLoc3models2019}.

\begin{table}[!t]
\centering
\caption{Hierarchical model performance evaluation.}
\begin{tabular}{cccccc}
\hline
Model & B-ACC & F-ACC & ACC & 2D-MDE & 3D-MDE \\
\hline
\hline
H-MLP & 99.85\% & 99.55\% & 99.55\% & 5.84m & 6.20m\\
\hline
MLP & 99.85\%  & 99.55\%  & 99.40\% & 7.69m & 8.12m \\
\hline
\end{tabular}
\label{tab:mlp}
\end{table}

\begin{table}[!t]
\centering
\caption{Benchmarck on UJIndoorLoc validation data.}
\begin{tabular}{cccccc}
\hline
Model & B-ACC(\%) & F-ACC(\%)  & 2D-MDE(m)  \\
\hline
\hline
Proposed H-MLP & 99.90 & 94.87 & 8.80  \\
\hline
HADNN\cite{auxiliary2022} & 100 & 93.15  &  14.93   \\
\hline
CNNLoc\cite{CNNLoc3models2019} & 99.27 & 96.03 & 18.10   \\
\hline
Scalable DNN\cite{scalableDNNLocIndoor2017} & 99.82 & 91.21 & 9.29   \\
\hline
\end{tabular}
\label{tab:bcmk}
\end{table}

\subsection{Evaluation of the FL framework}

\begin{figure}[!t]
    \centering
    \includegraphics[scale=0.8]{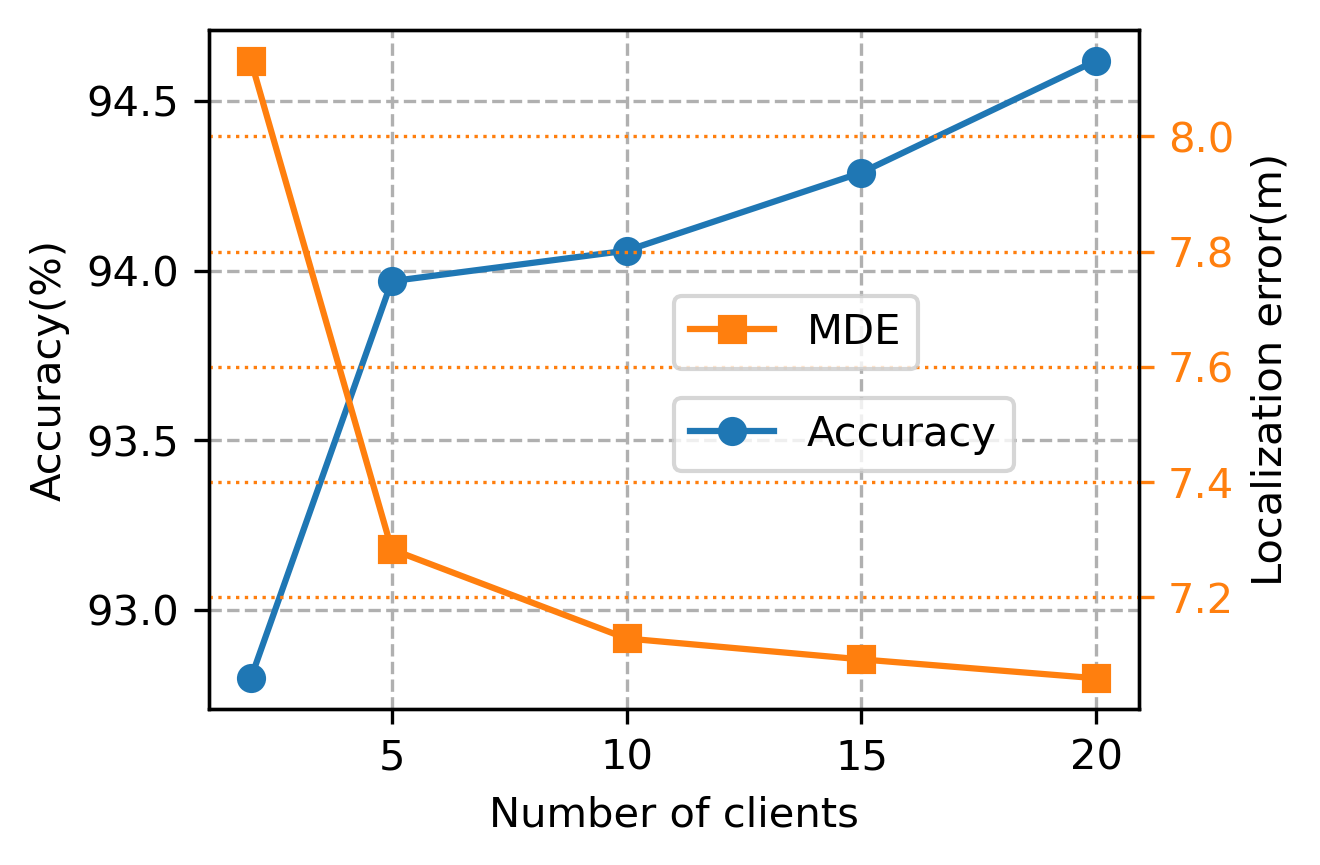}
    \caption{Training performance with increasing participants.}
    \label{fig:fl-scale}
\end{figure}

In this part, we train the previously defined model (H-MLP) in a federated manner. To do so, we keep the same configurations for the model and we set up an FL system whose configuration is shown in TABLE~\ref{tab:sim-config}. The learning curve and the accuracy of the FL system are respectively depicted in Fig.~\ref{fig:fl-loss} and Fig.~\ref{fig:fl-accuracy}, where it can be seen that the model starts to converge after 20 communication rounds. Therefore, as shown in Fig.~\ref{fig:fl-central}, FL clients can collaboratively train a unified model and achieve a near-performance characteristic as of the central training since for instance, the localization error increases by only 7.69\%.
Furthermore, we investigate the scalability property of the proposed FL framework as well as the communication budget with a focus on the uplink. Note that the communication resources optimization of the FL framework and clients selection strategy are out of the scope of this work and left for future work.
As such, Fig.~\ref{fig:fl-scale} depicts the localization performance with an increasing number of clients participating in federated training. It is observed that the success rate increases with the number of clients, and the localization error gets better. Indeed, with more clients participating, more data are seen by the model which consequently keeps improving.

\begin{figure}[!t]
    \centering
    \includegraphics[scale=0.8]{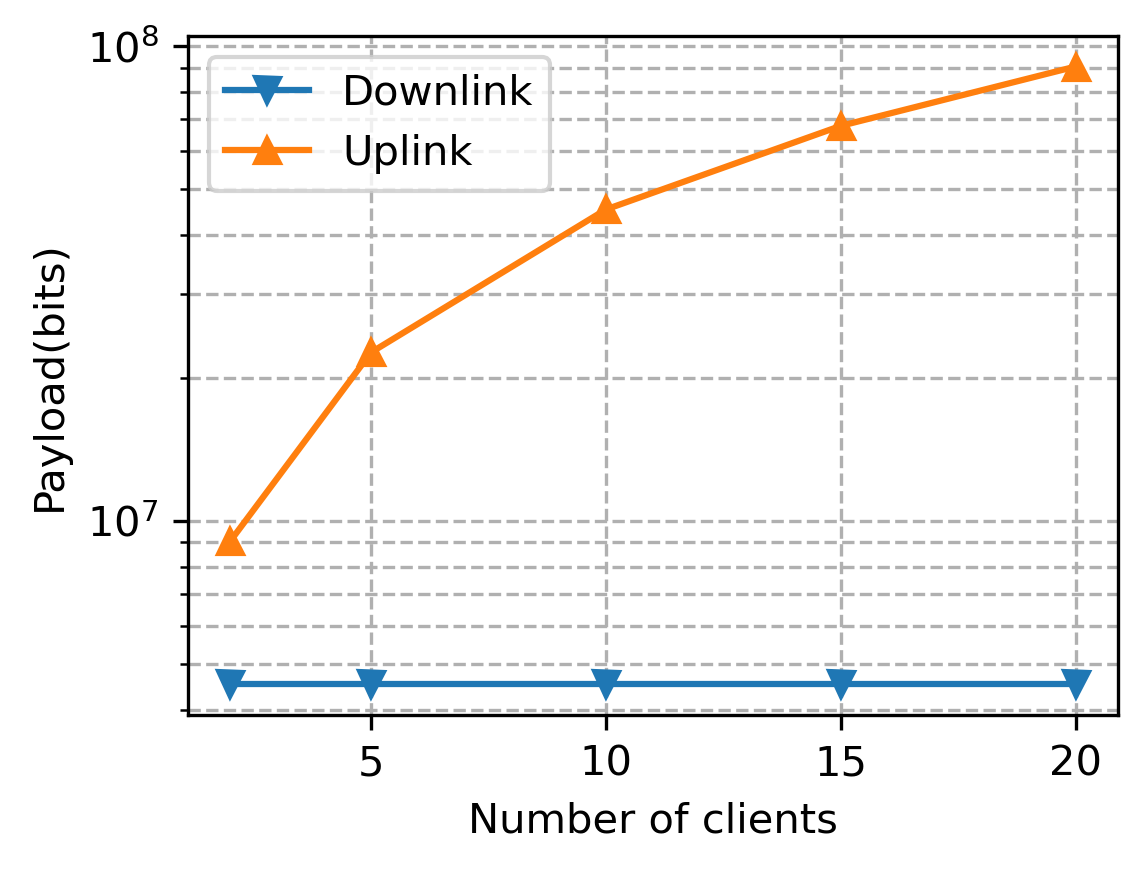}
    \caption{Communication load vs. number of participants.}
    \label{fig:fl-com}
\end{figure}

However, these improvements in the localization performance come at the price of an increase in the communication load, especially in the uplink transmission as shown in Fig.~\ref{fig:fl-com} where we considered a bit resolution R=32 so that the number of transmitted bits by a single device is given by $W\times R$ where $W$ is the total number of the model parameters. Indeed, the downlink remains quasi-unchanged no matter the number of participating clients since it is the same model that is being broadcast each time.
 On the other hand, on the uplink, the more participants that share the spectrum equally, the more bandwidth is required.

\begin{table}[!t]
\caption{FL Simulation settings. }
\centering
    \begin{tabular}{ |l|l|l| }
   \hline
 \textbf{Parameter} & \textbf{Description} & \textbf{Value}  \\
 \hline
 Optimizer &  Model optimizer & Adam  \\
 \hline
 $\eta$ & Learning rate & 0.0005    \\
\hline
$\beta_1, \beta_2$  & Exponential decay rates & $0.1, 0.99$  \\
\hline
$\mathcal{C}$ & Number of clients & 5\\  
\hline
$B$ & Batch size & 64\\  
\hline
$E$ & Number of epochs per local iteration & 10\\
\hline
$\mathcal{R}$ & Communication rounds & 100\\  
\hline
\end{tabular}
\label{tab:sim-config}
\end{table}
\vspace{-0.1in}

\section{Conclusion}
\label{sec:conclusion}
In this work, we have first of all pointed out the prominence of exploiting the hierarchical property of indoor localization and then
presented a federated learning framework to implement the proposed model.
The performance analysis has shown that our proposed
model can achieve an improvement in the localization error with a reduction of up to 24.06\% in addition to the good compromise with the building and floor hit rates which are respectively 99.90\% and 94.87\%.
Furthermore, we have demonstrated that at the cost of an increase in the communication load, the FL-based localization system performance improves when more participants are added.
However, regarding the limited communication resources of IoT devices participating in the training, there are some limitations in terms of the DNN architectures that can be transmitted over the wireless network and also in terms of the number of participants. Therefore, further investigation on communication resources management in the FL setting is needed in future works.



\bibliographystyle{IEEEtran}
\bibliography{references}









\end{document}